\documentclass{article}

\usepackage[htt]{hyphenat}
\usepackage{amsmath}
\usepackage{amssymb}
\usepackage{authblk}
\usepackage{graphicx}
\usepackage{hyperref}
\usepackage{natbib}

\usepackage[frozencache,cachedir=.]{minted}

\setminted{fontsize=\footnotesize}

\begin{document}
\def\thetitle{HEIR:\@ A Universal Compiler for Homomorphic Encryption}
\title{\thetitle}

\author[1]{Asra Ali\thanks{Authorship is listed alphabetically by surname.}}
\author[2]{Jaeho Choi\thanks{Author contributed while at Korea Advanced Institute of Science \& Technology (KAIST)}}
\author[1]{Bryant Gipson}
\author[1]{Shruthi Gorantala}
\author[1]{Jeremy Kun}
\author[1,3]{Wouter Legiest}
\author[4]{Lawrence Lim}
\author[5]{Alexander Viand}
\author[6]{Meron Zerihun Demissie}
\author[7]{Hongren Zheng}
\affil[1]{Google}
\affil[2]{Moreh}
\affil[3]{Katholieke Universiteit Leuven}
\affil[4]{University of California, Santa Barbara}
\affil[5]{Intel}
\affil[6]{University of Michigan, Ann Arbor}
\affil[7]{Tsinghua University}

\maketitle

\begin{abstract}
This work presents Homomorphic Encryption Intermediate Representation (HEIR),
a unified approach to building homomorphic encryption (HE) compilers.
HEIR aims to support all mainstream techniques in homomorphic encryption,
integrate with all major software libraries and hardware accelerators,
and advance the field by providing a platform for research and benchmarking.
Built on the MLIR compiler framework,
HEIR introduces HE-specific abstraction layers
at which existing optimizations and new research ideas
may be easily implemented.
Although many HE optimization techniques have been proposed,
it remains difficult to combine or compare them effectively.
HEIR provides a means to effectively explore the space of HE optimizations.
HEIR addresses the entire HE stack and includes support for various frontends,
including Python.
The contribution of this work includes:
(1) We introduce HEIR as a framework for building HE compilers.
(2) We validate HEIR's design by porting a large fraction of the HE literature to HEIR,
and we argue that HEIR can tackle more complicated and diverse programs than prior literature.
(3) We provide evidence that HEIR is emerging as the de facto HE compiler
for academic research and industry development.
\end{abstract}

\section{Introduction}
Homomorphic encryption (HE) is a cryptographic tool
that allows one to run programs directly on encrypted data.
Once described as the ``holy grail'' of cryptography~\cite{rivest1978data},
decades of research have produced
numerous homomorphic encryption (HE) schemes,
hardware accelerators,
and HE programming techniques.
Collectively, these advances have resulted
in the deployment of HE-based systems
at major companies like Apple, Microsoft, and Google~\cite{wally-search,chen2018labeled,Ion2020}.
However, to the best of the authors' knowledge,
all publicly-described production systems using HE
are implemented with hand-optimized HE subroutines,
required teams of cryptography experts to build,
and only utilize comparatively simple HE circuits.
By contrast, recent HE research literature has focused
on the private evaluation of machine learning
models such as deep convolutional networks~\cite{Lee22,Lee22-pp,Podschwadt2020,Park2023}.
These research papers often also hand-optimize the translation
of the network to an HE circuit.
The increasing breadth and ambitions of HE research and practice
raise the natural need for compiler support.
While there have been many research initiatives and in-house efforts within private companies towards building HE compilers~\cite{SP:ViaJatHit21},
all prior work has been specialized
for particular tasks or types of programs.
Many research HE compilers are built purely to demonstrate
just one or two optimization techniques.

We introduce Homomorphic Encryption Intermediate Representation (HEIR),
a general purpose HE compiler.\footnote{HEIR is open source and under active development at \url{https://github.com/google/heir}}
As Section~\ref{sec:overview} explains in detail,
HEIR defines layers of abstraction for HE computations
at many levels, as well as a series of compiler passes
that lower from higher layers of abstraction to lower layers.
Each layer is designed to make particular optimizations easy to implement,
such as ciphertext data layout (aka ``packing''),
parameter selection,
and arithmetization.
HEIR is also a platform for HE research.
By providing an existing syntax, parsers, intermediate representations (IRs),
standard compiler passes, and a multitude of backend targets,
HEIR allows researchers to focus on their novel optimization.
By additionally incorporating existing HE techniques,
one can use HEIR as a comparison and benchmarking platform.
Indeed, works including~\cite{MalikCircuitOU} used HEIR as a foundation,
which we describe in Section~\ref{sec:standard}.

This paper is organized as follows.
Section~\ref{sec:highlevel}
provides a high-level overview of homomorphic encryption
and how the constraints of HE create
challenges in compiler design.
Section~\ref{sec:comparison} compares HEIR with prior research.
Section~\ref{sec:overview} provides an overview of HEIR's
design and architecture.
Section~\ref{sec:optimization} highlights some of the notable
optimization passes and analyses implemented in HEIR.\@
Section~\ref{sec:pipeline} describes how the passes
are organized into sample pipelines, and demonstrates
those pipelines via examples.
Section~\ref{sec:standard} describes the ways in which
HEIR is becoming the de facto standard for HE compiler
research and development.

\section{High-level overview of HE}\label{sec:highlevel}

Homomorphic encryption schemes can be split into roughly two categories
based on the types of programs they are well-suited for.
The first class of schemes,
which we call \emph{scalar} HE\footnote{This is often called
\emph{boolean} HE in the literature in reference to the early
schemes that were designed to encrypt individual bits.
Modern analogues can often encrypt small integers.},
is identified by the property
that ciphertexts typically encrypt a single scalar value.
This class includes schemes like DM~\cite{EC:DucMic15} and CGGI~\cite{JC:CGGI20}.
The second class of schemes,
which we call \emph{vector} HE\footnote{This is also called \emph{arithmetic} HE in the literature},
are by contrast characterized by ciphertexts encrypting
large vectors of data.
In vector HE, each ciphertext encrypts a large vector,
typically of size $2^{10}$ to $2^{17}$.
By contrast with traditional CPU vector registers,
these ``ciphertext registers'' are very large,
yet they have a strictly limited set of operations:
elementwise addition and multiplication,
and a cyclic rotation\footnote{Generally it is not precisely a cyclic rotation, but a
restricted class of permutations that is close to a cyclic rotation semantically.
The most common kind of permutation views a vector register as a $(2 \times N/2)$-sized matrix
and supports cyclic rotations of the columns and cyclic rotation of (i.e., swapping) the rows.}
operation in lieu of an in-register shuffle.
This class of schemes includes BGV~\cite{ITCS:BraGenVai12}, BFV~\cite{EPRINT:FanVer12}, and CKKS~\cite{AC:CKKS17}.
By contrast, scalar HE schemes typically
feature dedicated operations for applying a univariate lookup table to an encrypted input.

All modern HE schemes are based on variants of
the learning with errors (LWE) problem, in particular ring LWE (RLWE).
As such, all of the HE schemes considered in this paper
hide the message in random noise.
Applying homomorphic operations generally increases the noise,
with multiplication increasing noise significantly faster than addition.
Eventually, the noise grows too large to support further operations,
and so programs must end or else schemes must support
a technique called \emph{bootstrapping}
to reset the noise to a fixed level independent of the ciphertext.
Scalar HE schemes typically have fast bootstrapping techniques
(on the order of millisecond on CPU),
while vector schemes have slow bootstrapping operations
(on the order of seconds on CPU).
However, due to the large vector sizes used in vector schemes,
the amortized cost of bootstrapping can be less
in vector schemes than scalar schemes.
Some schemes also support the concept of \emph{programmable bootstrapping},
which allows one to evaluate a static, fixed-size univariate lookup table
at the same time as reducing noise~\cite{JC:CGGI20, EPRINT:AleKimPol24}.
For a more detailed description of the major HE schemes
and their operations, see~\cite{Marcolla22}.

Many papers have been written about techniques to manually
craft HE programs that solve specific problems using a specific scheme~\cite{EPRINT:LDSTV25,Lee22}.
Writing HE programs to target scalar HE schemes
typically involves low-level circuit optimization,
reducing the bit width of the intermediate computations,
and designing clever ways to shift computation
into the programmable bootstrapping lookup tables.
Meanwhile, writing programs for vector HE schemes
typically focuses on carefully packing data
within the ciphertexts---in conjunction with
carefully chosen kernels for operations
like matrix-vector multiplication---so as to
minimize the number of elementwise multiplications
and ciphertext rotations
(the most expensive operations besides bootstrap).

In all HE schemes,
functions that cannot be expressed as a polynomial require special care to implement.
In scalar HE schemes,
non-polynomial computations are typically achieved via lookup tables,
while vector HE schemes use polynomial approximations,
along with various analytical and empirical methods
to reduce the degree of the approximations~\cite{Lee22-minimax,Tong24}.
Such polynomial approximations also introduce
error into the computation,
and so aggressive degree reduction is only suitable for some applications,
such as machine learning,
where performance is routinely traded off against accuracy loss.

Given the above,
designing a compiler for HE
involves many novel challenges,
including circuit synthesis in a new circuit model posed by scalar HE schemes
and programmable bootstrapping,
data layout problems posed by the large ``ciphertext register'' model
of vector HE,
and approximations which impact program correctness.
The methods to solve these problems
interact and conflict with cryptosystem parameter selection,
making it challenging to holistically trade off program latency,
memory usage,
ciphertext expansion and key material size (i.e., bandwidth costs),
and security parameters.
Furthermore, a good compiler must also select
the right optimizations for a particular hardware target,
and many upcoming HE accelerators
only support a subset of the many tricks used in designing HE programs.

\section{Comparison to prior literature}\label{sec:comparison}

A plethora of HE compilers and HE-specific program optimizations have been proposed.
While early works struggled to offer meaningful speedups~\cite{SP:ViaJatHit21},
recent research has demonstrated
that automated optimizations can offer significant performance improvements~\cite{USENIX:VJHH23},
in some cases beating hand-optimized programs~\cite{USENIX:CLKLJK24, Lee2022-nh, USENIX:LCKLK23}.

Existing work falls into two broad categories: general purpose and application specific HE compilers.
General-purpose compilers face a more difficult task,
as they need to support a wide range of different possible input programs.
While many proposed techniques show promise when considered in the context of smaller programs,
they often fail to scale to larger programs
as they rely often on expensive searches through exponentially large design spaces.
For example, works like the Google Transpiler~\cite{GoogleTranspiler} and Porcupine~\cite{Cowan21Porcupine}
that rely on synthesis tend to take hours to compile programs of modest size.
Other works like HECO~\cite{USENIX:VJHH23} and HECATE~\cite{HECATE}
effectively require full loop unrolling,
which quickly becomes impractical for deep loop nests.
Still other works are limited to specific subsets of the overall compilation pipeline,
such as HeaNN-mlir~\cite{HEANN-mlir} and Cinnamon~\cite{Cinnamon}.
All application-specific HE compilers in the literature specialize to neural network inference.
These compilers achieve success more readily
due to a highly constrained program structure,
lack of control flow,
and limited primitives~\cite{Krastev2024-is,Aharoni2023,Orion,SEALion,ngraphHE,Cinnamon}.

An increasing number of recent works in this area
have utilized the Multi-Level Intermediate Representation (MLIR) compiler framework~\cite{Lattner2021-km}.
MLIR provides a flexible and extensible infrastructure for building domain specific compilers,
and has been applied to a variety of domains including machine learning~\cite{StableHLO},
quantum computing~\cite{MLIR-Quantum}, and hardware synthesis~\cite{CIRCT}.
MLIR provides a rich set of features
for defining new intermediate representations (IRs);
a variety of built-in IRs for basic arithmetic, control-flow, and memory management;
existing optimizations for operating on the built-in abstractions;
and set of generic optimizations, such as common subexpression elimination,
that can operate on arbitrary IRs including those defined by extensions like HEIR.\@
HE compiler works that utilize MLIR include~\cite{USENIX:VJHH23,HECATE,Concrete,Bian2024,HEANN-mlir}.

Despite a trend towards MLIR-based compilers,
the FHE software ecosystem has remained highly fragmented and incompatible.
The vast majority of works only consider a single scheme,
target a single backend,
and implementation decisions are often tied to the chosen backend's limitations.
For example, works that target SEAL~\cite{SEAL} tend not to support bootstrapping
because SEAL does not implement it.
This even extends to tools that use the same language and framework.
Of all the prior works that use MLIR, few reuse each other's components,
because the specificity of the work bleeds into the IR definitions
to the point that they are not reusable.
While there has been some limited re-use
(e.g., Bian et al.~\cite{NDSS:BZZMSJGL24} re-use some of the passes from HECO~\cite{USENIX:VJHH23}),
it is generally limited to hard forks to manually reconcile incompatibilities.
Ultimately, many of the proposed optimizations in the HE literature are not reusable across tools.
This fragmentation wastes significant effort in the research community by reimplementing common components,
including HE-agnostic routines like common subexpression elimination
and common HE-aware methods
like Halevi-Shoup matrix-vector multiplication~\cite{Halevi2014-hn}.
%
HEIR comes with many of these techniques already implemented,
allowing a researcher working with HEIR
to focus on their novel optimization
rather than reinvent the wheel.

HEIR aims to provide a suite of benchmarks and example programs
that compile to all supported backend targets
(accelerators and library APIs).
This allows researchers to easily perform
more comprehensive testing of their methods,
whereas today most HE papers test
against a single backend or hardware target.
More generally, HEIR aims to be a universal platform that can support all techniques in HE.\@
Where existing FHE compilers focus on a subset of techniques, or specific HE schemes and backends;
HEIR can incorporate different methods and constraints flexibly.
One consequence of this charter
is that HEIR augments and complements other HE toolchains
that focus on specific techniques or subsets of the HE stack.
For example, as mentioned in Section~\ref{sec:codegen},
HEIR has integrations with many existing hardware accelerators;
HEIR provides compilation pipelines that provide high-level
program analysis, ciphertext packing, and parameter selection,
and then exports the IR to the hardware vendor's
lower-level toolchain to manage scheduling and assembly.\footnote{
  HEIR ultimately aims to incorporate similar low-level passes
  directly, but currently these toolchains are the protected intellectual
  property of the companies building them.}
Similarly, works like Cinnamon~\cite{Cinnamon} have explicitly cited HEIR
as potentially targeting the Cinnamon ISA.\@
We posit that HEIR's generality,
as well as the project's explicit charter to support research,
will prove to be a force multiplier for the HE research community.

\section{Overview of HEIR design}\label{sec:overview}

\begin{figure}[!tb]
  \centering
  \includegraphics[width=0.4\columnwidth]{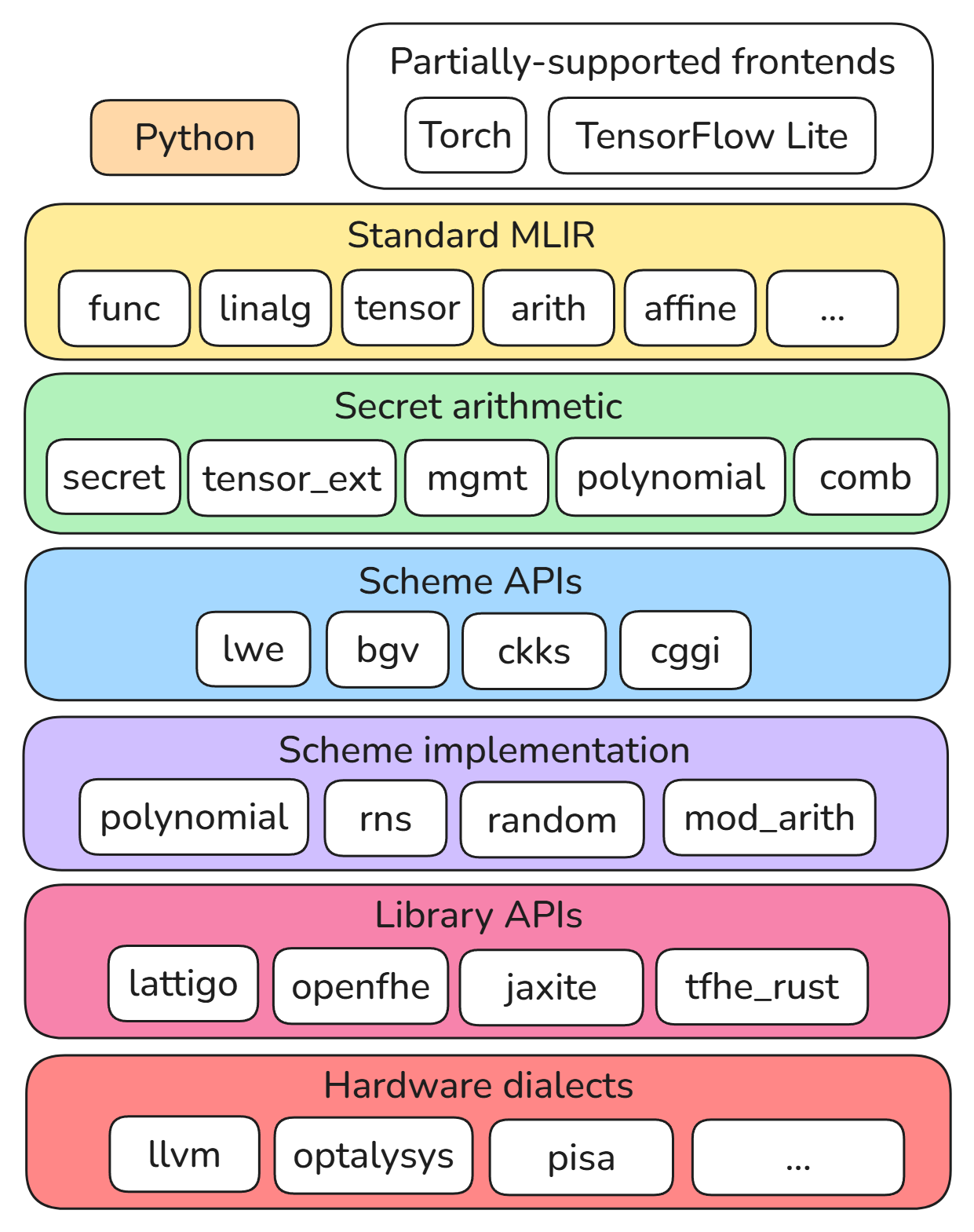}
  \caption{
    A summary of the dialects in HEIR, grouped roughly by abstraction layer.
    The ``secret arithmetic'' layer represents scheme-agnostic HE programs
    in the scalar or vector abstract FHE models.
    ``Scheme APIs'' represents high-level scheme-specific HE operations
    and materialized ciphertext types.
    ``Scheme implementation'' represents the modular arithmetic and ring operations
    implementing each scheme operation.
    ``Library APIs'' contain dialects that mirror external APIs to support
    code generation against them.
    ``Hardware dialects'' represent dialects that target specific hardware platforms,
    including LLVM for CPU.\@
    Compilation flows downward in the diagram, though some compilation pipelines
    may skip layers or exit early, e.g., to generate code against library APIs.
  }\label{fig:dialectdiagram}
\end{figure}

While HEIR does not definitively solve the challenges posed in Section~\ref{sec:highlevel},
by bringing together many techniques from the HE literature into one system,
it enables systematic comparisons, benchmarking, and experimentation
to further the state of the art in homomorphic encryption.

HEIR is built using the Multi-Level Intermediate Representation (MLIR) framework~\cite{Lattner2021-km}.
This provides HEIR with standardized language syntax
and passes related to arithmetic, tensors, control flow,
and much more,
as well as compiler pass orchestration and debugging tools.
HEIR extends MLIR by defining new intermediate representations (IR),
called \emph{dialects},
for HE operations at many relevant layers of abstraction.
This includes scheme-agnostic programs and circuits that operate on secret data,
the ``ciphertext register'' computational model
that abstracts over vector HE schemes,
scheme-specific dialects for all major HE schemes,
low-level polynomial, residue number system, and modular integer arithmetic,
crypto library APIs like OpenFHE and Lattigo,
and hardware ISAs.
These new dialects introduce additional types, such as LWE ciphertexts;
operations, such as ciphertext-ciphertext multiplication or bootstrapping;
and attributes, which are data attached to types and operations
identifying HE-relevant quantities
like ciphertext noise and cryptosystem parameters.
This architecture is summarized in Figure~\ref{fig:dialectdiagram}.
By way of example, Figure~\ref{fig:dotprod1} shows a function
computing a dot product, expressed in standard MLIR,
where the private inputs are marked with the \texttt{secret.secret} annotation.
Then Figure~\ref{fig:dotprod2} shows the same function
after being lowered through one of HEIR's default vector HE compilation pipelines
to the \texttt{lattigo} dialect, which is a HEIR dialect
designed to mirror the Lattigo library API~\cite{Lattigo}.

Next we provide details of the dialects at the different layers of abstraction.
These will be referenced in Section~\ref{sec:optimization} when discussing
the abstraction layer various optimization passes operate on.

\begin{figure}[!tb]
\begin{minted}{mlir}
func.func @dot_product(
    %arg0: tensor<8xi16> {secret.secret},
    %arg1: tensor<8xi16> {secret.secret}) -> i16 {
  %c0 = arith.constant 0 : index
  %c0_si16 = arith.constant 0 : i16
  %0 = affine.for %arg2 = 0 to 8 iter_args(%iter = %c0_si16) -> (i16) {
    %1 = tensor.extract %arg0[%arg2] : tensor<8xi16>
    %2 = tensor.extract %arg1[%arg2] : tensor<8xi16>
    %3 = arith.muli %1, %2 : i16
    %4 = arith.addi %iter, %3 : i16
    affine.yield %4 : i16
  }
  return %0 : i16
}
\end{minted}
\caption{An example of a dot product computation
  in vanilla MLIR, with \texttt{secret.secret} annotations
  for what will compile to encrypted inputs.
}\label{fig:dotprod1}
\end{figure}

\begin{figure}[!tb]
\begin{minted}{mlir}
!ct = !lattigo.rlwe.ciphertext
!encoder = !lattigo.encoder
!evaluator = !lattigo.evaluator
!param = !lattigo.parameter
func.func @dot_product(
    %evaluator: !evaluator, param: !param, %encoder: !encoder,
    %ct: !ct, %ct_0: !ct) -> !ct {
  %c7 = arith.constant 7
  %c1_i16 = arith.constant 1
  %cst = arith.constant dense<0>
  %inserted = tensor.insert %c1_i16 into %cst[%c7]
  %ct_1 = lattigo.mul %evaluator, %ct, %ct_0, %ct_0
  %ct_2 = lattigo.relinearize %evaluator, %ct_1, %ct_0
  %ct_3 = lattigo.rotate_columns %evaluator, %ct_2, %ct {offset = 4}
  %ct_4 = lattigo.add %evaluator, %ct_2, %ct_3, %ct_0
  %ct_5 = lattigo.rotate_columns %evaluator, %ct_4, %ct {offset = 2}
  %ct_6 = lattigo.add %evaluator, %ct_4, %ct_5, %ct_0
  %ct_7 = lattigo.rotate_columns %evaluator, %ct_6, %ct {offset = 1}
  %ct_8 = lattigo.add %evaluator, %ct_6, %ct_7, %ct_0
  %ct_9 = lattigo.rescale %evaluator, %ct_8, %ct_0
  %pt = lattigo.new_plaintext %param
  %pt_10 = lattigo.encode %encoder, %inserted, %pt {scale = 5350}
  %ct_11 = lattigo.mul %evaluator, %ct_9, %pt_10, %ct_0
  %ct_12 = lattigo.rotate_columns %evaluator, %ct_11, %ct_0 {offset = 7}
  %ct_13 = lattigo.rescale %evaluator, %ct_12, %ct_0
  return %ct_13
}
\end{minted}
\caption{The IR from Figure~\ref{fig:dotprod1} after optimizing
and lowering to the \texttt{lattigo} dialect.
The loop has been replaced by a network of rotations.
At this point, the \texttt{heir-translate} tool converts this IR
to Go code using the Lattigo API.
Note the MLIR types and generated client code have been omitted for brevity.}\label{fig:dotprod2}
\end{figure}

\subsection{Scheme-agnostic secret computation}

HE operations and types introduce constraints
that make it incompatible with some existing MLIR infrastructure.
For example, the MLIR \texttt{arith} dialect enforces operation inputs
to be typed as integers, floats, or containers of these types,
and so it cannot be reused for types like ciphertexts defined by HEIR.\@
In order to reuse as much of this existing infrastructure as possible,
we define the \texttt{secret} dialect
that wraps standard MLIR computations and types,
while allowing HEIR to reuse all existing MLIR passes
before lowering a program to use ciphertext types directly.

The \texttt{secret.secret} type wraps an arbitrary type
to denote a Static Single Assignment (SSA) value that is semantically encrypted.
The \texttt{secret.generic} op defines an enclosed basic block
that lifts cleartext computations to operate on secret values.
The operands of a \texttt{generic} specify which secret values
should be made available to the computation as cleartext values.
The \texttt{secret.yield} op converts the cleartext results back to secrets.
An example is given in Figure~\ref{fig:secretgeneric},
which is equivalent to Figure~\ref{fig:dotprod1}
after applying the \texttt{wrap-generic} pass.
Inside a \texttt{generic} op,
any upstream MLIR transformation may be applied
as if there were no \texttt{secret} types.

\begin{figure}[!tb]
\begin{minted}{mlir}
func.func @dot_product(
    %arg0: !secret.secret<tensor<8xi16>>,
    %arg1: !secret.secret<tensor<8xi16>>) -> !secret.secret<i16> {
  %c0_i16 = arith.constant 0 : i16
  %0 = secret.generic ins(
      %arg0, %arg1 :
      !secret.secret<tensor<8xi16>>, !secret.secret<tensor<8xi16>>) {
  ^bb0(%arg2: tensor<8xi16>, %arg3: tensor<8xi16>):
    %1 = affine.for %arg4 = 0 to 8
         iter_args(%arg5 = %c0_i16) -> (i16) {
      %extracted = tensor.extract %arg2[%arg4] : tensor<8xi16>
      %extracted_0 = tensor.extract %arg3[%arg4] : tensor<8xi16>
      %2 = arith.muli %extracted, %extracted_0 : i16
      %3 = arith.addi %arg5, %2 : i16
      affine.yield %3 : i16
    }
    secret.yield %1 : i16
  } -> !secret.secret<i16>
  return %0 : !secret.secret<i16>
}
\end{minted}
\caption{An example of a dot product computation
wrapped in \texttt{secret.generic}.
Inside the generic, secret values are converted
to their underlying cleartext values
and a computation is performed.
Within the \texttt{generic},
any upstream MLIR pass can be applied.}\label{fig:secretgeneric}
\end{figure}

The \texttt{secret} dialect also
supports a dataflow analysis called \texttt{SecretnessAnalysis}.
It can identify whether an operation is ciphertext-ciphertext type or
ciphertext-plaintext type. Plaintext-plaintext type operations will be lifted
out of the \texttt{secret.generic} block by canonicalizer.

Secret-typed values are later converted to ciphertext types,
while \texttt{secret.generic} operations are converted to scheme-specific operations,
but at this stage the HE scheme and parameters have not yet been chosen.
Once the scheme is known, HEIR splits \texttt{generic} operations apart
so that each \texttt{generic} contains exactly one (non-\texttt{yield}) cleartext op.
Later passes can then match on these simplified operations
to determine how to convert them to ciphertext-ciphertext
or ciphertext-plaintext operations in a particular scheme.

\subsection{Scheme-agnostic HE computation}

In addition to upstream MLIR passes,
there are many transformations
that are easiest to express on cleartext computations
or in a manner that generalizes over some subset of HE schemes.
We defer the discussion of the optimization details to Section~\ref{sec:optimization}.

For example, the \texttt{comb} dialect defines operations
for standard boolean circuits and lookup tables.
Compilation flows targeting scalar HE schemes like CGGI
may use a circuit synthesizer
to generate \texttt{comb} programs
that can then be lowered to CGGI.

The \texttt{mgmt} dialect corresponds to
ciphertext management operations,
including \texttt{mod\_reduce}, \texttt{relinearize}, and \texttt{bootstrap}.
An initial pass inserts \texttt{mgmt} operations,
and then analyses and optimizations can be applied,
such as noise analysis, parameter selection,
and optimal placement of management operations.
While \texttt{mgmt} operations are not fully scheme-agnostic,
many can be reused across schemes, and their semantics are interpreted
differently by compiler passes that are configured in a scheme-specific way.

The \texttt{tensor\_ext} dialect
corresponds to the operations in the
abstract vector HE programming model,
including operations like \texttt{tensor\_ext.rotate}
that performs a cyclic rotation of a tensor,
as well as operations to manage ciphertext data layout.
The optimizations expressed in \texttt{tensor\_ext}
are best expressed in terms of the ``vector register'' model.

\subsection{Scheme-specific dialects}

Next HEIR defines dialects, types, and operations
that are specific to particular HE schemes.
This includes scheme-specific variations in message types,
encodings, and modulus-chain management.

The \texttt{lwe} dialect defines common types and operations across all schemes,
such as LWE and RLWE ciphertext types,
secret key types,
encoding/decoding and encryption/decryption.
This dialect provides a pass
that add encryption and decryption helper functions
that are exposed as part of the public API
for the compiled result.

The \texttt{bgv}, \texttt{ckks}, and \texttt{cggi}
dialects contain scheme-specific operations for
the BGV/BFV, CKKS, and CGGI schemes, respectively.
Operations in these dialects may overlap in functionality,
but must be distinct operations
if their implementations differ significantly.
The BFV scheme is implemented
as a special case of BGV with different encoding parameters.

\subsection{Code generation dialects}\label{sec:codegen}

Two primary purposes of HEIR
include targeting hardware accelerators
and supporting comparisons against existing HE implementations.
HEIR supports these with infrastructure for generating code
at varying degrees of abstraction.
We divide this into a two-pronged approach:
scheme-level code generation for library APIs,
and low-level hardware dialects for particular accelerators.

To support code generation
for existing cryptosystem libraries, HEIR introduces
dialects that are intended to mirror external HE library APIs.
These include
\texttt{openfhe} for OpenFHE's~\cite{OpenFHE}
vector HE implementations (CKKS, BGV, and BFV),
\texttt{lattigo} for Lattigo~\cite{Lattigo},
\texttt{tfhe\_rust} and \texttt{tfhe\_rust\_bool} for TFHE-rs~\cite{TFHE-rs},
and \texttt{jaxite} and \texttt{jaxite\_word}
for CGGI and CKKS subsets of Jaxite, a TPU implementation of these schemes~\cite{Jaxite}.
HEIR provides a \texttt{heir-translate} tool
that generates C++, Go, Rust, or Python code for
each of these dialects.

Generating source code has two added benefits.
First, many HE accelerator toolchains today
use a mirror of an existing cryptosystem library API as their frontend.
This makes it easy to initially integrate hardware with HEIR
via a library intermediary.
Second, it allows researchers to fairly compare HEIR-generated programs
against existing programs written for a library API.\@

The second, preferred approach
is to implement the lower-level hardware optimizations as HEIR compiler passes.
HEIR has ongoing collaborations with hardware designers to achieve this
(cf. Section~\ref{sec:standard}),
and we provide the example of the \texttt{pisa} dialect, a hardware-specific variant
of HEIR's general \texttt{polynomial} dialect (cf. Section~\ref{sec:mathdialects}).
The \texttt{pisa} dialect models the Polynomial Instruction Set Architecture (PISA) used
to target Intel's HERACLES accelerator~\cite{PISA,Cammarota2022-ru}.

\subsection{Low level math dialects}\label{sec:mathdialects}

To support hardware code generation,
HEIR must implement cryptosystem primitives as compiler passes.
To support this, HEIR defines four dialects.
The \texttt{random} dialect is used to define
random distributions and random number sampling operations for encryption.
The \texttt{mod\_arith} dialect contains
types and operations for modular arithmetic in an integer ring $\mathbb{Z}/n\mathbb{Z}$.
The \texttt{rns} dialect defines a type for values
represented in a residue number system (RNS).
Finally, the \texttt{polynomial} dialect
represents arithmetic on single-variable polynomials in a quotient ring.
This is primarily intended to support
RNS polynomial types, which are the central data type for many vector HE schemes,
though nontrivial polynomial quotient rings
are also required for the implementation of scalar HE schemes.

The \texttt{polynomial} dialect includes typical ring ops (add, mul, modulus switching),
as well as performance-oriented operations representing implementation details
of HE schemes and other polynomial operations (e.g., a number theoretic transform,
used to implement polynomial multiplication).
The ability to specify any quotient polynomial and any coefficient type
ensures the \texttt{polynomial} dialect can support
a broad variety of HE parameter choices,
such as the variants of vector HE schemes that use nontrivial hypercube rotations.
The \texttt{polynomial} dialect is also reused
for polynomial approximation, as it defines
a convenient static representation of polynomials
and an \texttt{eval} op representing polynomial evaluation
on a compatible data type.

\subsection{Frontends}

\begin{figure}[!tb]
\begin{minted}{python}
import numpy as np
from heir import compile
from heir.mlir import I16, Secret, Tensor

@compile()  # defaults to scheme="bgv", OpenFHE backend
def func(x: Secret[Tensor[1024, I16]], y: Secret[Tensor[1024, I16]]):
  result = x
  for i in range(10):
    result = result + y
  return result * result

x = [v for v in range(1024)]
y = [2 * v for v in range(1024)]
print("Cleartext: " + func.original(np.array(x), np.array(y)))
print("Compiled: " + func(x, y))
\end{minted}
\caption{An example Python program compiled via HEIR's Python frontend.}\label{fig:pythonfrontend}
\end{figure}

HEIR ships with a Python frontend, demonstrated in Figure~\ref{fig:pythonfrontend}.
The frontend converts Python bytecode---paired with special type annotations
to denote what is secret---to  HEIR's \texttt{secret} dialect.
The frontend performs basic type inference,
supports tensor-typed arguments,
loops, and control flow.
The subset of acceptable Python includes
restrictions typical of HE programs
(e.g., loops require iteration bounds to not depend on secrets).
For supported library API backends,
the Python frontend also coordinates
generating the library API code,
further compiling it to machine code,
and loading the shared object files
into Python with a sensible API,
so that users need not read or write MLIR directly.

HEIR also partially supports StableHLO~\cite{StableHLO} as a frontend
for converting trained machine learning inference models to homomorphic encryption.
In particular, StableHLO's \texttt{legalize-to-linalg} pass can be used
to convert StableHLO programs to MLIR's \texttt{linalg} dialect,
at which point one must annotate the entry function
to denote which inputs are secret.
Work is ongoing to provide full support for the StableHLO API in HEIR.
Notably, systems like PyTorch support exporting to StableHLO.

In theory, any frontend which can be converted to standard MLIR
can be adapted for compatibility with HEIR.
This opens the possibility for future frontends in
C++ (via ClangIR~\cite{ClangIR}),
ONNX (via \texttt{onnx-mlir}~\cite{ONNX-mlir}),

\section{Optimization passes}\label{sec:optimization}

In this section we outline optimization passes of interest in HEIR.
Note that HEIR inherits a variety of traditional compiler optimizations
from the upstream MLIR project,
such as dead code elimination,
common subexpression elimination,
and constant propagation.
It is worthwhile to pause to consider how beneficial this is
from the perspective of a researcher using HEIR as a platform (cf. Section~\ref{sec:standard}):
at the time of this writing, MLIR has 237 built-in passes,
and many more optimizations expressed as peephole rewrite patterns
under the ``canonicalization''  and ``folding'' umbrellas.
HEIR, meanwhile, introduces another 88 passes,
along with its own collection of further canonicalization and folding patterns.
With that understanding,
this section focuses on a selection of key passes in HEIR
that are primarily relevant for homomorphic encryption.

\subsection{Ciphertext management}

All HE schemes must insert additional operations to ``manage'' ciphertexts.
Scalar HE schemes typically use bootstrapping to manage noise.
BGV employes \textit{modulus switching}, a division with rounding to reduce noise growth;
CKKS repurposes it as \textit{rescaling} to manage message encoding precision.
vector HE schemes typically use \textit{relinearization} to reduce ciphertext size
after the tensor product step in multiplication.
Each of these operations comes with choices about when to execute them,
performance costs, and noise impacts associated with their deferral.

For vector HE schemes,
HEIR defines passes
\texttt{secret-insert-mgmt-\{bgv,bfv,ckks\}}
at the \texttt{secret} dialect level
to insert baseline ciphertext management operations for BGV/BFV and CKKS.\@
For BGV and CKKS, the pass inserts relinearizations after each multiplication,
modulus switching before multiplications (not including the first multiplication),
and proper scaling factor adjustment when there is a mismatch,
following the conventions of OpenFHE and~\cite{Kim2021-rn}.
There are also pass options to control whether to insert modulus switching right after multiplications,
or to insert modulus switching before each multiplication.
For BFV, the pass only handles relinearizations as there is no modulus switching.
For CKKS, the pass also implements
an initial bootstrap placement via a greedy algorithm.
HEIR does not support BGV/BFV bootstrapping as currently there is no backend support for it.
For scalar HE schemes, HEIR replaces lookup table operations from the circuit synthesis engine
with programmable-bootstrap operations, and does not further attempt to reduce bootstrapping.

The management of scaling factor is implemented in three parts. These
\texttt{secret-insert-mgmt} passes only use placeholder operations to signal
there is a possible scaling factor mismatch. Only after the parameter selection
pass do we know the concrete value of the scaling factor, then
\texttt{populate-scale-\{bgv,ckks\}} passes are run to replace the placeholder
operations with concrete scaling factor adjustments.

Next, HEIR includes passes to improve
upon the initial placements.
First HEIR defines an \texttt{optimize-relinearization} pass,
which optimizes the placement of \texttt{mgmt.relinearize} operations.
It defines a mixed-integer linear program (ILP)
adapted from the optimal bootstrap placement ILP of Paindavoine and Vialla~\cite{Paindavoine2016-rn}.
The ILP is implemented using Google's \texttt{or-tools} library~\cite{Perron2011-cf}
with SCIP~\cite{Bestuzheva2023-sh} as the underlying ILP solver.

\subsection{Parameter selection and noise modeling}\label{sec:paramselect}

Parameter selection is a critical step in any HE application,
as it has a strong impact on both security and performance.
The primary advantage that an HE compiler has when doing parameter selection
is that it has access to the entire program.
By analyzing what operations are performed,
and the connection between them,
a compiler can produce tighter parameters
in contrast to a general-purpose library API\footnote{Some libraries like OpenFHE
are, at the time of this writing, working on providing
parameter selection APIs that take some information about
the circuit into account, cf.~\cite{cryptoeprint:2024/203}.
However, these APIs are inevitably more coarse-grained
than the compiler's view of the entire program.}
or user-chosen empirical parameters.

HEIR contains noise models for BGV and BFV,
and uses these models to guide parameter selection.
While HEIR also has noise models available for CKKS,
they require explicit bounds on the range of values
of the underlying messages, which requires additional
user-specific inputs.

Two of the BGV noise models track the evolution of bounds on infinity norm
of the ciphertext coefficient embedding, given in~\cite{Kim2021-rn}.
One of these models assumes worst-case expansion during multiplication (producing rather loose bounds),
while the other uses an empirical average-case expansion factor (producing bounds with margin of around 10 bits).
A third BGV noise model, from~\cite{Murphy2024-zf}, tracks the variance of coefficients
using the Central Limit argument.
The last model produces much tighter bounds,
often with a margin of a few bits, but may also underestimate.
Additional implemented noise models include the BFV model of~\cite{EPRINT:BMCM23}
and the BGV model of~\cite{AFRICACRYPT:MMLGA23} and its BFV adaption.
All models support public-key and secret-key variants.

HEIR provides parameter selection
for BGV that analyzes the maximal noise for each RNS level
and picks tight prime moduli that could ``reset'' the noise by modulus switching
for ciphertexts at that level.
The knowledge of noise specific to the program dramatically simplifies more complex, program-oblivious
parameter selection routines in the literature like~\cite{AC:KimPolZuc21} and runtime
noise management like~\cite{C:HalSho14,EPRINT:HalSho20}
while also providing tighter noise bounds.

For BFV, HEIR analyzes the overall noise growth and selects a tight moduli chain
consisting of 60 bit primes. For CKKS, HEIR currently asks the user to provide
the bit width of the first modulus and scaling moduli (i.e. the scaling factor),
and notifies the user
when the first modulus may be too small using range analysis.

HEIR also generates other parameters like the auxiliary
moduli used by hybrid key switching~\cite{C:GenHalSma12} and the ring dimension by
looking up the 128-bit security parameter table from~\cite{cryptoeprint:2024/463},
assuming uniform ternary secrets being used in backends.

To break the chicken-and-egg problem in noise analysis and parameter generation,
HEIR first uses the conservative choice
of all 60-bit moduli to initialize the noise model,
then uses a forward-propagated noise analysis for the above generation.

\subsection{Layout selection}

Layout selection, also called ciphertext packing,
is the process of jointly deciding how to lay out
application data among ciphertexts
while also choosing efficient kernels for key operations,
such as matrix multiplication.

To represent a layout,
HEIR uses MLIR's concept of a \emph{quasi-affine map},
abbreviated to ``affine map'' in this paper.
An affine map is an $m$-tuple of expressions

\[
(d_1, \dots, d_n) \to (e_1(d_1, \dots, d_n), \dots, e_m(d_1, \dots, d_n))
\]

where each $e_i$ is an affine formula of the input dimensions
with the additional allowed operations
\texttt{ceildiv}, \texttt{floordiv}, and \texttt{mod},
which implement division rounding to the ceiling,
division rounding to the floor, and integer remainder, respsectively,
and all of these operations require
an integer literal for the divisor or modulus.

An affine map is interpreted as a ciphertext layout
by mapping a tensor of shape $s = (s_1, \dots, s_n)$
to a tensor of ciphertexts of shape $(e_1, \dots, e_{m-1})$.
The ciphertexts in this tensor have the same fixed number of slots
(usually a power of two).
The input tensor's entry at index $d = (d_1, \dots, d_n)$
is stored in slot $e_m(d)$ of ciphertext $(e_1(d), \dots, e_{m-1}(d))$.
For example, a Halevi-Shoup-style matrix-vector multiplication~\cite{Halevi2014-hn}
requires packing an $n \times m$ matrix in ciphertexts of size $N$,
and this can be expressed as the affine map
\texttt{(i, j) -> ((j - i) mod n, (j - ((j - i) mod n)) mod N)}.\footnote{
  Note this mapping maps from the data matrix to the ciphertext index and slot.
  A more familiar mapping in the literature maps from ciphertext slots to
  the data matrix entry, namely \texttt{(i, j) -> (j mod n, (i+j) mod m)}. The map
  given here is the inverse of that map.}

HEIR's initial layout selection algorithm
is split into a few passes, following the basic outline in the Fhelipe compiler~\cite{Krastev2024-is}.
First, the \texttt{layout-propagation} pass
applies a pass that sets default row-major layouts
for all input tensors in a program,
and applies a forward propagation of the layouts through the program.
When an operation is encountered
that has one or more incompatible layouts for its operands,
\texttt{layout-propagation} inserts (potentially expensive) layout conversion operations.
Second, a \texttt{layout-optimizer} pass applies a backward propagation
attempting to hoist layout conversion operations through operations
toward the beginning of the program.
At each such operation, the decision on whether to hoist,
and which of many possible conversions to hoist,
is determined greedily according to a cost model:
the cost of layout conversions, the operation's implementation
for the particular layouts, and opportunistic mergings
of adjacent conversion operations are considered to get a net
increase in the cost of the hoist,
and then only hoists with negative net cost are hoisted.

Finally, a \texttt{convert-to-ciphertext-semantics}
pass materializes the chosen layout while also implementing
operation kernels that are sensitive to layout details.
During the process of choosing and implementing layouts,
HEIR does not yet convert a tensor data type to \emph{explicit} ciphertext types
(i.e., a \texttt{lwe.ciphertext} type),
opting instead for a semantic interpretation of an MLIR \texttt{tensor} type
as having \emph{ciphertext semantics}
by treating its trailing dimension as representing ciphertext slots.
This is required partly due to type constraints in MLIR's \texttt{linalg}
and \texttt{tensor} dialects,
and partly because some optimizations like the initial parameter selection routine of Section~\ref{sec:paramselect}
can still be performed more easily on ``ciphertext semantic'' tensors than ciphertexts,
as ciphertext types cannot fully be specified without parameters available.

One important example of such a transformation
is the process of implementing a layout conversion operation.
A layout conversion may apply an arbitrary permutation
to the slots of a ciphertext,
but this must be implemented in the vector HE model
of rotations, addition, and multiplication
(notably using plaintext-ciphertext multiplications
to mask out particular subsets of ciphertext slots before rotating).
In the HE literature, finding an efficient implementation of a layout conversion
is often called the \emph{shift network} problem.
HEIR provides an \emph{implement-shift-network} pass
that implements the graph coloring approach
of Vos-Vos-Erkin~\cite{Vos2022-dq}.

\subsection{Polynomial approximation}

HEIR includes a polynomial approximation solver
based on the Caratheodory-Fejer (CF) method~\cite{Van_Deun2011-uq}.
The CF approximation is based on the eigenvalue analysis
of a Hankel matrix of Chebyshev coefficients.
CF polynomial approximation differs
from the Remez methods used in many HE systems~\cite{Lee21Remez}.
The Remez algorithm is notoriously difficult,
while the CF algorithm is much simpler to implement.
And, as per~\cite{Van_Deun2011-uq}, for most functions
the CF approximation is indistinguishable
(up to machine precision) from
the best approximation as produced by Remez.
HEIR currently requires input annotations specifying
the degree and intervals of approximation.

After an op has been approximated, HEIR replaces
it with a \texttt{polynomial.eval} op,
representing the evaluation of a static polynomial
with an SSA-valued input.
This is further lowered
to the underlying arithmetic operations using the methods of Paterson-Stockmeyer~\cite{Paterson1973-am}
as a means to trade-off the costs
of ciphertext multiplication and addition.

\subsection{Other optimizations}

This section describes additional passes provided by HEIR
that do not fit into the larger categories above.

HEIR includes a variety of passes
that attempt to convert a program built using HE-unfriendly constructs
to HE-friendly counterparts.
For example, \texttt{convert-if-to-select}
converts if statements to the multiplexed analogue \texttt{arith.select},
and \texttt{convert-secret-for-to-static-for} converts for loops
with data-dependent bounds
to loops that are independent of their bounds
(though users must annotate the loops with static upper bounds).
HEIR also defines an \texttt{operation-balancer} pass
that attempts to rebalance trees of arithmetic operations
so as to provide lower multiplicative depth,
adapting a pass from the EVA compiler~\cite{Chowdhary2021-ow}.

For scalar HE schemes,
HEIR provides an \texttt{yosys-optimizer} pass
that invokes the Yosys optimization suite~\cite{Wolf2013yosys}
on the body of each \texttt{secret.generic} op,
replacing the body with the synthesized circuit.
There are two notable aspects of this pass.
First, it supports generating circuits
with arbitrary sets of gates.
We include configuration corresponding to standard boolean gates
as well as lookup-tables of various bit widths,
which are used to model programmable bootstrapping in some CGGI backends.
Second, the pass provides the ability to partially unroll loop nests
and optimize only the innermost body of the result.
This allows the compiler to trade off the optimality of the generated program
with the compiler runtime and memory usage
incurred by running Yosys on large, fully-unrolled programs.

At the \texttt{tensor\_ext} layer,
HEIR has ported the complete functionality of
the HECO compiler~\cite{USENIX:VJHH23}, including
passes like \texttt{insert-rotate} that implements HECO's slot target insertion heuristic,
and \texttt{rotate-and-reduce} that identifies opportunities to
convert linear reductions of tensors (e.g., summing all elements of a ciphertext)
to equivalent reductions that take advantage of slot alignment
to use only a logarithmic number of rotations.

HEIR uses a single \texttt{straight-line-vectorizer} pass that applies batching
or vectorization on groups of compatible operations.
It runs a graph analysis on the IR to identify what can be executed concurrently,
and supports bounds the batch size.
It is applied at the scheme level to
efficiently schedule operations for maximum throughput on target devices.
Specifically, it is used for multi-threading on CPU,
parallel execution on single- and multi-FPGA backends like FPT~\cite{Van_Beirendonck2023-ni},
and batching for TPUs in Jaxite~\cite{Jaxite}. This may also
be used at lower layers, for example, \texttt{polynomial}, to automatically
detect limb-level parallelization in RNS variants of RLWE schemes.

Finally, HEIR provides the \texttt{alloc-to-inplace} pass
which converts a program written against a library API
to use in-place API calls where possible,
reducing unnecessary memory allocations.

\subsection{Integration passes}

HEIR also provides passes that do not perform
optimizations, but are necessary for using HE programs
in practice.

HEIR provides the \texttt{lwe-add-client-interface} pass
which inserts client encryption and decryption routines
that handle plaintext encoding and type conversions necessary
to integrate the compiled function with non-HE client code.
Similar passes are provided for each library API
to handle setup and configuration of crytographic contexts,
such as what automorphism key material to generate
and what features of the library API must be enabled.

HEIR also provides a \texttt{plaintext} backend.
This is a CPU backend target that operates
in the plaintext domain of an HE scheme,
without actually encrypting anything.
Beyond being useful as a reference implementation
to identify bugs in compilation passes,
it also aids in manually selecting BGV/BFV plaintext modulus and CKKS scaling factor,
as doing this requires prior knowledge of intermediate value bounds.

Finally, HEIR provides a pass \texttt{insert-debug-handler-calls}
to insert an external function call after each homomorphic operation,
while also threading through the secret key.
This allows one to add arbitrary intermediate computations,
such as logging and message introspection,
to a HEIR-compiled program.
This is used by HEIR developers
to compare actual ciphertext noise to noise model bounds,
and to identify the operation at which a message decrypts incorrectly
while debugging a faulty optimization pass.

\section{Pass pipelines}\label{sec:pipeline}


In this section we describe pass pipelines
that combine the individual optimization passes above with dialect lowerings
to produce complete HE compilation pipelines.
We demonstrate these pipelines with an example program.

\subsection{\texttt{mlir-to-cggi} and AES-128 Transciphering}\label{sec:aes}

\begin{figure}[!tb]
\begin{minted}{mlir}
#map = affine_map<(d0, d1) -> (d0, d1)>
#map1 = affine_map<(d0, d1) -> (d1)>
#map2 = affine_map<(d0, d1) -> (d0)>
module {
  func.func private @mul_gf256_2(
      %arg0: i8 {secret.secret}, %arg1: i8) -> i8 {
    %c27_i8 = arith.constant 27 : i8
    %c0_i8 = arith.constant 0 : i8
    %c1_i8 = arith.constant 1 : i8
    %0:2 = affine.for %arg2 = 0 to 2
        iter_args(%arg3 = %arg0, %arg4 = %c0_i8) -> (i8, i8) {
      %1 = arith.index_cast %arg2 : index to i8
      %2 = arith.shrsi %arg1, %1 : i8
      %3 = arith.andi %2, %c1_i8 : i8
      %4 = arith.trunci %3 : i8 to i1
      %5 = arith.xori %arg4, %arg3 : i8
      %6 = arith.select %4, %5, %arg4 : i8
      %7 = arith.cmpi sle, %arg3, %c0_i8 : i8
      %8 = arith.shli %arg3, %c1_i8 : i8
      %9 = arith.xori %8, %c27_i8 : i8
      %10 = arith.select %7, %9, %8 : i8
      affine.yield %10, %6 : i8, i8
    }
    return %0#1 : i8
  }
  func.func private @mix_single_column(
      %arg0: tensor<4xi8> {secret.secret}) -> tensor<4xi8> {
    %cst = arith.constant dense<
      [[2, 3, 1, 1],
       [1, 2, 3, 1],
       [1, 1, 2, 3],
       [3, 1, 1, 2]]> : tensor<4x4xi8>
    %cst_0 = arith.constant dense<0> : tensor<4xi8>
    %0 = linalg.generic
      {indexing_maps = [#map, #map1, #map2],
       iterator_types = ["parallel", "reduction"]
      } ins(%cst, %arg0 : tensor<4x4xi8>, tensor<4xi8>)
        outs(%cst_0 : tensor<4xi8>) {
    ^bb0(%in: i8, %in_1: i8, %out: i8):
      %1 = func.call @mul_gf256_2(%in_1, %in) : (i8, i8) -> i8
      %2 = arith.xori %out, %1 : i8
      linalg.yield %2 : i8
    } -> tensor<4xi8>
    return %0 : tensor<4xi8>
  }
}
\end{minted}
\caption{MixColumns implemented in standard MLIR.}\label{fig:mlirmixcolumns}
\end{figure}

\begin{figure}[!tb]
\begin{minted}{mlir}
    <elided>
    %pt = lwe.encode %false
    %ct_5 = lwe.trivial_encrypt %pt
    %ct_6 = cggi.lut3 %ct_5, %ct_3, %ct_4 {lookup_table = 1 : ui8}
    %ct_7 = memref.load %arg0[%c0, %c1]
    %ct_8 = memref.load %arg0[%c0, %c6]
    %ct_9 = cggi.lut3 %ct_5, %ct_7, %ct_8 {lookup_table = 1 : ui8}
    %ct_10 = cggi.lut3 %ct_9, %ct_6, %ct_2 {lookup_table = 128 : ui8}
    <elided>
\end{minted}
\caption{MixColumns lowered to CGGI operations.}\label{fig:cggimixcolumns}
\end{figure}

\begin{figure}[!tb]
\begin{minted}{rust}
pub fn encrypt_block(
    server_key: &ServerKey,
    block: &SecretBlock,
    key: &[SecretBlock; 11],
    n_rounds: usize
  ) -> SecretBlock {
    let mut block = add_round_key::add_round_key(
        &server_key, &block, &key[0]);
    for i in 1..(n_rounds + 1) {
        block = sub_bytes_par::sub_bytes(&server_key, &block);
        block = shift_rows::shift_rows(&block);
        block = mix_columns::mix_columns(&server_key, &block);
        block = add_round_key::add_round_key(
            &server_key, &block, &key[i]);
    }
    block = sub_bytes_par::sub_bytes(&server_key, &block);
    block = shift_rows::shift_rows(&block);
    block = mix_columns::mix_columns(&server_key, &block);
    block = add_round_key::add_round_key(
         &server_key, &block, &key[n_rounds + 1]);
    block
}
\end{minted}
\caption{A driver combining the AES subroutines.}\label{fig:aesdriver}
\end{figure}

The HEIR pipeline \texttt{mlir-to-cggi} converts secret-annotated MLIR
to the \texttt{cggi} dialect.
First, the pipeline automatically removes and data-dependent conditions
and unrolls loops with static bounds.
The pipeline can also be configured to partially unroll loops,
treating the inner body of a loop nest as a subcircuit.
Next, the pipeline uses the Yosys~\cite{Wolf2013yosys} circuit synthesis tool
to create and optimize a circuit representation of the program.
This step can be configured to emit
a circuit that uses a specific subset of boolean gates,
or else a circuit that uses lookup-table evaluations in place of gates.
The latter is useful for programmable bootstrapping.
The pipeline then lowers the gate operations to the \texttt{cggi} dialect,
using leveled LWE operations to prepare gate inputs
and programmable bootstrapping to evaluate the gates.

HEIR generates backend code for CGGI on CPU through the
\texttt{tfhe-rs}~\cite{TFHE-rs} library in Rust, Belfort's FPGA in Rust, and
Google's TPU through the \texttt{jaxite} Python library. Like mentioned above,
HEIR can vectorize operations and emit batched or parallelized code for these
backends with the same pass.

To demonstrate this pipeline, we outline an implementation of AES in HEIR.\@
Advanced Encryption Standard (AES) is a commonly used symmetric block cipher in
which several rounds consisting of 4 operations are applied to a message to
obtain an encrypted message. The 128-bits AES algorithm uses 10 rounds and a key
and block size of 128 bits. AES is considered a poor candidate for
arithmetic-based HE schemes, due to the non-linearity of the round operations.
Each of the four round operations can be written in standard MLIR.\@ For
example, \texttt{MixColumns}, treats a column of four bytes as a polynomial and
performs a polynomial multiplication with a constant polynomial
$3x^3 + x^2 + x + 2$.
Polynomial multiplication is manifested as a matrix multiplication with
the \texttt{linalg.generic} op, using the \texttt{mul\_gf256\_2} function to
multiply and the XOR operation to add in $GF(256)$ (cf.
Figure~\ref{fig:mlirmixcolumns}).

The main kernel \texttt{mix\_column} applies \texttt{mix\_single\_column} for each
row of the block. The pipeline will unroll the loops of the multiplication
kernel and inline the fixed polynomial's constants. The circuit synthesis tools
will find an optimized representation of the multiplication and lowers the
circuit IR to the corresponding CGGI operations
(Figure~\ref{fig:cggimixcolumns}). A user-provided driver
(Figure~\ref{fig:aesdriver}) combines all four compiled kernels into one
execution.

\subsection{\texttt{mlir-to-ckks} and matrix-vector multiplication}

HEIR's \texttt{mlir-to-ckks} pipeline similarly starts from
secret-annotated MLIR.\@ But unlike the CGGI pipeline in Section~\ref{sec:aes},
it does not unroll loops or convert the program to a circuit.
Instead, the pipeline applies polynomial approximation,
layout selection, and linear-algebra operation kernels,
followed by ciphertext management, noise analysis, and parameter selection routines.\footnote{
As noted in Section~\ref{sec:paramselect}, CKKS parameter selection
is not completely automated, but analogous pipelines are defined for BGV and BFV.}
Finally, the pipeline lowers the program to the \texttt{ckks} dialect,
from which point one can choose a desired backend to continue compilation and code generation.

Figure~\ref{fig:matvecexample} shows an example IR
that composes two feed-forward linear layers
of a hypothetical neural network, with ReLU activations.
The pipeline which identifies the \texttt{arith.maximumf} op having a constant operand,
and converts it to a \texttt{polynomial.eval} op
(cf. Figure~\ref{fig:matvecexample-poly}),
and then implements the `\texttt{eval}' op using Paterson-Stockmeyer.
Next, the layout selection passes run,
selecting both ``alignment'' (padding and repetition of the cleartext tensor)
and layout maps for the matrices and vectors
which are then propagated forward the IR and annotated on op results.
(cf. Figure~\ref{fig:matvecexample-layout}).
Then the pipeline implements \texttt{matvec} using the Halevi-Shoup method
and inserts \texttt{assign\_layout} operations to pack the plaintext matrices
appropriately (cf. Figure~\ref{fig:matvecexample-kernel}).
Next, the ciphertext management passes, noise analysis,
and parameter selection passes run, until the IR is finally
lowered to CKKS scheme operations and emitted as OpenFHE code (Figure~\ref{fig:matvecexample-openfhe}).

\begin{figure}[!tb]
\begin{minted}{mlir}
func.func @matvec(
    %arg0: tensor<16xf32> {secret.secret}) -> tensor<16xf32> {
  %matrix1 = arith.constant dense<1.0> : tensor<16x16xf32>
  %matrix2 = arith.constant dense<2.0> : tensor<16x16xf32>
  %cst_1 = arith.constant dense<0.0> : tensor<16xf32>
  %0 = linalg.matvec
    ins(%matrix1, %arg0 : tensor<16x16xf32>, tensor<16xf32>)
    outs(%cst_1 : tensor<16xf32>) -> tensor<16xf32>
  %1 = arith.maximumf %0, %cst_1 : tensor<16xf32>
  %2 = linalg.matvec
    ins(%matrix2, %1 : tensor<16x16xf32>, tensor<16xf32>)
    outs(%cst_1 : tensor<16xf32>) -> tensor<16xf32>
  %3 = arith.maximumf %2, %cst_1 : tensor<16xf32>
  return %3 : tensor<16xf32>
}
\end{minted}
\caption{Matrix-vector multiplication followed by a ReLU, twice.}\label{fig:matvecexample}
\end{figure}

\begin{figure}[!tb]
\begin{minted}{mlir}
#ring_f64 = #polynomial.ring<coefficientType = f64>
!poly = !polynomial.polynomial<ring = #ring_f64>
#poly1 = #polynomial<typed_float_polynomial<
      0.0383 + <elided> + -1.574E-16x**5> : !poly>
module attributes {backend.openfhe, scheme.ckks} {
  func.func @matvec(...) {
    <elided>
    %1 = linalg.matvec ...
    %2 = polynomial.eval #poly1, %1 : tensor<16xf32>
    %3 = linalg.matvec ...
    %4 = polynomial.eval #poly1, %3 : tensor<16xf32>
    <elided>
  }
}
\end{minted}
\caption{Polynomials approximating ReLUs}\label{fig:matvecexample-poly}
\end{figure}

\begin{figure}[!tb]
\begin{minted}{mlir}
#alignment = #tensor_ext.alignment<in = [16], out = [1024]>
#alignment1 = #tensor_ext.alignment<in = [16, 16], out = [16, 16]>
#vec = #tensor_ext.layout<map = (d0) -> (d0), alignment = #alignment>
#diag = #tensor_ext.layout<map = (d0, d1) -> (
    (d1 - d0) mod 16,
    (d1 - ((d1 - d0) mod 16)) mod 1024
  ), alignment = #alignment1>
module attributes {backend.openfhe, scheme.ckks} {
  func.func @matvec(...)
    %0 = secret.generic ins(%arg0)
        attrs = {__argattrs = [{tensor_ext.layout = #vec}],
                 __resattrs = [{tensor_ext.layout = #vec}]} {
    ^body(%input0):
      %1 = tensor_ext.assign_layout %matrix1 {layout = #diag}
      %2 = tensor_ext.assign_layout %cst_5 {layout = #vec}
      %3 = linalg.matvec {layout = #vec} ...
      %4 = arith.mulf %3, %3 {layout = #vec}
      // ... elided ReLU implementation operations ...
      %21 = arith.addf %20, %12 {layout = #vec}
      %22 = tensor_ext.assign_layout %matrix2 {layout = #diag}
      %23 = tensor_ext.assign_layout %cst_5 {layout = #vec}
      %24 = linalg.matvec {layout = #vec} ins(%22, %21
      %25 = arith.mulf %24, %24 {layout = #vec}
      // ... elided ReLU implementation operations ...
      %42 = arith.addf %41, %33 {layout = #vec}
      secret.yield %42
    }
    return %0
  }
}
\end{minted}
\caption{The matrix-vector example after layout selection.
In this example, ciphertexts have 1024 slots.
Some MLIR syntax has been elided for clarity.}\label{fig:matvecexample-layout}
\end{figure}

\begin{figure}[!tb]
\begin{minted}{mlir}
<elided>
%c0_i64 = arith.constant 0
%4 = tensor_ext.rotate %input0, %c0_i64
%extracted_slice = tensor.extract_slice %1[0, 0] [1, 1024] [1, 1]
%5 = arith.mulf %4, %extracted_slice
%6 = arith.addf %3, %5
%c1_i64 = arith.constant 1
%7 = tensor_ext.rotate %input0, %c1_i64
%extracted_slice_73 = tensor.extract_slice %1[1, 0] [1, 1024] [1, 1]
%8 = arith.mulf %7, %extracted_slice_73
%9 = arith.addf %6, %8
<elided>
\end{minted}
\caption{A subset of the IR for the diagonalized kernel
for matrix-vector multiplication.}\label{fig:matvecexample-kernel}
\end{figure}

\begin{figure}[!tb]
\begin{minted}{cpp}
CiphertextT matvec(CryptoContextT cc, CiphertextT ct) {
  ...
  const auto& pt2 = cc->MakeCKKSPackedPlaintext(v501_filled);
  const auto& ct5 = cc->EvalMult(ct4, pt2);
  const auto& ct6 = cc->EvalRotate(ct, 3);
  ...
  const auto& ct46 = cc->EvalAdd(ct41, ct45);
  const auto& ct47 = cc->EvalAdd(ct38, ct46);
  const auto& ct48 = cc->EvalMultNoRelin(ct47, ct47);
  const auto& ct49 = cc->Relinearize(ct48);
  ...
}

// client functions (implementations elided)
CiphertextT matvec__encrypt__arg0(
  CryptoContextT cc, std::vector<float> v0, PublicKeyT pk);
std::vector<float> matvec__decrypt__result0(
  CryptoContextT cc, CiphertextT ct, PrivateKeyT sk);
CryptoContextT matvec__generate_crypto_context();
CryptoContextT matvec__configure_crypto_context(
  CryptoContextT cc, PrivateKeyT sk);
\end{minted}
\caption{A sample of the emitted OpenFHE code.}\label{fig:matvecexample-openfhe}
\end{figure}

\section{HEIR as a standard for HE research and practice}\label{sec:standard}

In this section we provide evidence that HEIR is becoming
a standard for HE compiler research and industry development.

First, the HEIR team
has developed initial integrations
with many hardware accelerator backends,
using library API intermediaries (cf. Section~\ref{sec:codegen}).
These include:

\begin{itemize}

\item The Belfort FPGA and its FPT predecessor~\cite{Van_Beirendonck2023-ni}, which supports CGGI via the TFHE-rs API.

\item Optalysys's optical accelerator~\cite{kundu2021dawn}, which supports CGGI via the TFHE-rs API and a dedicated C API.

\item Google's TPU v6e (Trillium)~\cite{Trillium}, which supports CGGI and CKKS via the Jaxite API.

\item Niobium's BASALISC ASIC~\cite{Geelen2023-si}, which supports BGV and CKKS via the OpenFHE API.

\item Intel's HERACLES ASIC~\cite{Cammarota2022-ru}, which supports BGV and CKKS via the OpenFHE API.

\item Cornami uses HEIR internally to target their accelerator~\cite{Cornami}.

\end{itemize}

It is worth noting that HE hardware accelerators
operate at drastically different abstraction layers.
Accelerators like Belfort's
schedule high level HE operations as atomic units,
subsuming all implementation details
like number-theoretic transforms and key switching
into the hardware design.
Other accelerators like HERACLES operate at the lower level of
RNS polynomial arithmetic.
HEIR's design makes this otherwise large discrepancy trivial:
hardware targets that operate at a high level
can exit the compilation pipeline at scheme operations,
while others can lower further to \texttt{polynomial}
before generating code.
As a result,
supporting additional targets via library APIs,
such as GPUs or other accelerators,
would be a straightforward extension.

Second, the homomorphic encryption community is actively working
toward multiple kinds of standardization.
One relevant to HEIR is the
The Fully Homomorphic Encryption Technical Consortium on Hardware (FHETCH)~\cite{fhetch},
which, among other goals, aims to standardize a hardware abstraction layer
for upstream tools to target.
FHETCH identified HEIR as an excellent candidate for this standardization.
The Homomorphic Encryption Standardization organization~\cite{hes}
also established a working
group for compilers and accelerators,
which now focuses on HEIR.\@
FHE.org~\cite{fheorg} also maintains a discussion forum for HEIR
that facilitates community engagement and collaboration
with academic researchers who use HEIR.\@

Third, academics are already using HEIR as a platform for research.
COATL~\cite{MalikCircuitOU} implements a novel optimization for CGGI
by constructing lookup tables for programmable bootstrapping
alongside specially-chosen linear combinations combining ciphertexts
into lookup table inputs.
COATL was built on HEIR.\@
The authors utilize the tools that are conveniently packaged
with HEIR, notably Yosys~\cite{Wolf2013yosys} for the initial circuit booleanization
and Google's \texttt{or-tools} library~\cite{Perron2011-cf}
for their central optimization solver.
They also use Yosys as a baseline for comparing their optimized circuits
against a circuit synthesized without using
the linear-algebraic structure of the lookup tables.\footnote{
In addition to these tools, HEIR includes
the Eigen linear solver library~\cite{eigen} and
an FFT library~\cite{PocketFFT}, which are currently used
for implementing the CF polynomial approximation method.}
The authors are also aware of other in-progress research projects
built on HEIR that, at the time of writing, have yet to be published.

Homomorphic encryption research is increasingly becoming an interdisciplinary field of
theoretical cryptography, statistical analysis, hardware architecture,
distributed systems and compiler engineering. A major blocker to
practical research is lack of interoperable components and interdisciplinary expertise.
HEIR bridges this gap by providing an architecture that is designed
to be modular, reusable and interoperable. This framework allows
researchers to concentrate on their specific areas of expertise
while also smoothly integrating with the rest of the HE stack~\cite{CacmFheStack}.
As an increasingly large fraction of the HE literature
is implemented in HEIR, the benefits to researchers will compound.

\section{Acknowledgement}
This work was supported in part by the Horizon 2020 ERC Advanced Grant (101020005 Belfort)
and the CyberSecurity Research Flanders with reference number VOEWICS02.
Wouter Legiest is funded by FWO (Research Foundation - Flanders) as Strategic Basic (SB) PhD fellow (project number 1S57125N).
\includegraphics[width=0.2\columnwidth]{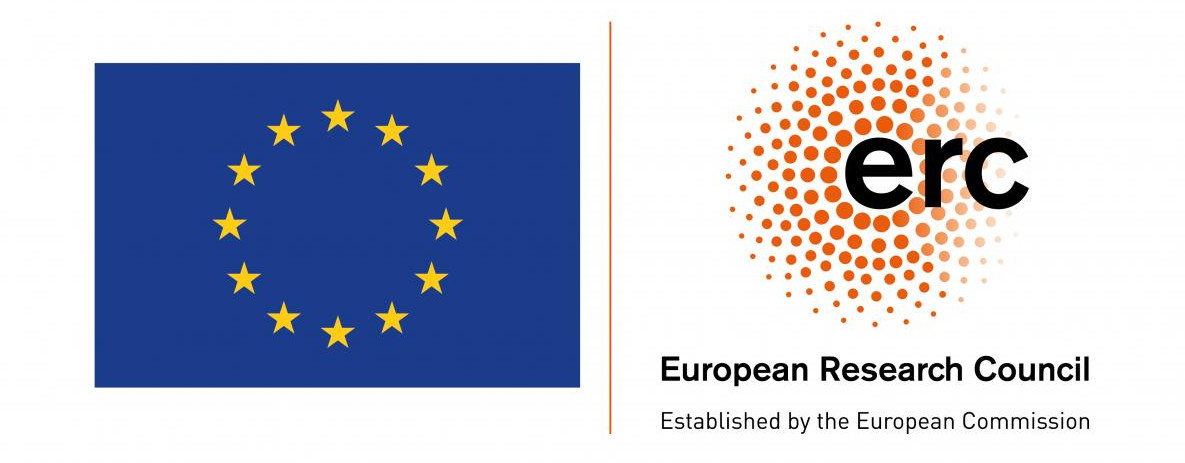}

\bibliographystyle{abbrv}
\bibliography{abbrev0, crypto, bib}

\end{document}